\documentclass[aps,pre,reprint,preprintnumbers,floatfix,nofootinbib]{revtex4-1}

\usepackage{amsmath}
\usepackage{amssymb}

\usepackage{graphicx}
\usepackage{subcaption}
\usepackage{epstopdf}
\usepackage{dcolumn}
\usepackage{bm}
\usepackage{marvosym}

\usepackage{color}
\definecolor{light-gray}{gray}{0.5}
\definecolor{blue}{rgb}{0.0,0.0,1.0}
\definecolor{green}{rgb}{0.0,0.5,0.0}
\definecolor{red}{rgb}{1.0,0.0,0.0}
\definecolor{cyan}{rgb}{0.0,0.75,0.75}
\definecolor{magenta}{rgb}{0.75,0.0,0.75}
\definecolor{yellow}{rgb}{0.75,0.75,0.0}

\newcommand{\avg}[1]{\langle{#1}\rangle}
\newcommand{\sdot}{\cdot}

\newcommand{\grad}{\bm \nabla}
\newcommand{\pd}{\partial}
\newcommand{\lrbig}[1]{\left( {#1} \right)}

\newcommand{\dd}{\mathrm{d}}

\begin{document}
\title{Self-organisation and non-linear dynamics in driven magnetohydrodynamic turbulent flows}
\author{V. Dallas}
\author{A. Alexakis}
\affiliation{Laboratoire de Physique Statistique, \'Ecole Normale Sup\'erieure, CNRS, Universit\'e Pierre et Mari\'e Curie, Universit\'e Paris Diderot, 24 rue Lhomond, 75005 Paris, France}

\begin{abstract}
Magnetohydrodynamic turbulent flows driven by random mechanical and electromagnetic external forces of zero helicities are investigated by means of direct numerical simulations. 
It is shown that despite the absence of helicities in the forcing, the system is attracted to self-organized helical states that exhibit laminar behaviour despite the large value of the Reynolds numbers examined. We demonstrate that the correlation time of the external forces is controlling the time spent
on these states, i.e. for short correlation times the system remains in the turbulent state while as the correlation time is increased the system spends more and more time in the self-organised states. As a result, time averaged statistics can significantly be affected by the time spent on these states.
These results have important theoretical implications for the understanding of the suppression of non-linearities in plasma fusion devises as well as in astrophysical observations.
\end{abstract}

\maketitle

\section{\label{sec:intro}Introduction}

Self-organisation in turbulent flows is the spontaneous creation of large-scale coherent structures out of a sea of homogeneous turbulence \cite{biskamp03}. In magnetohydrodynamic (MHD) theory, relaxation processes have been recognized to explain the evolution of electrically conducting fluids towards special states of self-organisation as a consequence of multiple conservation laws 
\cite{matthaeusetal12}. For homogeneous, incompressible, ideal 
MHD with zero mean magnetic field, there are three known quadratic conserved quantities:
the total energy      $E   \equiv E_u + E_b = \frac{1}{2} \avg{|\bm u|^2 + |\bm b|^2}_{_V}$, 
the magnetic helicity $H_b \equiv                         \avg{ \bm a \sdot \bm b   }_{_V}$ and 
the cross helicity    $H_c \equiv                         \avg{ \bm u \sdot \bm b   }_{_V}$.
Here, $\bm u$ is the velocity field, $\bm b \equiv \grad \times \bm a$ is the magnetic field where $\bm a$ is the solenoidal magnetic potential,
and $\avg{\sdot}_{_V}$ stands for spacial averages over the volume $V$.

The dissipative relaxation processes in decaying MHD turbulence are the Taylor relaxation \cite{chandrasekharwoltjer58,woltjer58,taylorJB74} 
and the dynamic alignment \cite{moffatt78,parker79}. These states could be derived analytically by minimizing an energy integral subject to 
some constraints \cite{biskamp03}. In detail, Taylor relaxation is associated with the decay of turbulence towards a minimum energy state 
under the constraint of finite $H_b$. The solution to this variational problem is a \textit{force-free} field, where $\bm u = 0$ and 
$\grad \times \bm b = \lambda' \bm b$ with the Lagrange multiplier $\lambda' \equiv \int \bm b^2_{min} \, dV / H_b$. 
On the other hand, the slow decay of $H_c$ in comparison to $E$ can lead to a minimum energy state while cross-helicity is conserved, 
where self-organisation occurs due to dynamic alignment between the velocity and the magnetic field, i.e. $\bm u = \pm \bm b$. 
This relaxed state is called \textit{Alfv\'enic state}. These relaxation processes can be seen as a \textit{selective decay} 
\cite{matthaeusmontgomery80} between the conserved quantities due to their very different rates of dissipation in turbulent flows.

The relaxation to the force-free state was offered as an explanation in reversed-field pinch plasma devices \cite{taylorJB74,taylorJB86} and it has also been used to estimate the energy release in coronal structures in connection to the problem of coronal heating \cite{parker83,heyvaertspriest84}.
Moreover, the Alfv\'enic states have been frequently observed in solar wind turbulence 
\cite{belcherdavis71,brunocarbone05}.
The growth of correlation between $\bm u$ and $\bm b$ in solar wind was conjectured to emerge dynamically from MHD turbulence \cite{dobrowolnyetal80} and this was verified in direct numerical simulations (DNS) of undriven turbulent flows \cite{pouquetetal86}. Moreover, Stribling and Matthaeus \cite{striblingmatthaeus91} have shown numerically that in a truncated model of three-dimensional decaying MHD turbulence the final states depend on the initial values of $H_b$ and $H_c$. In particular, they showed that the final state for strongly helical initial conditions is the force-free field, whereas for sufficient large initial alignment between $\bm u$ and $\bm b$ is the Alfv\'enic state.

In this paper, we show that self-organization due to force-free, Alfv\'enic and Beltrami (i.e. $\grad \times \bm u \propto \bm u$) states can occur in driven MHD turbulent flows from initial conditions and external forces of zero helicity. We also demonstrate that these states depend on the correlation time scale of the external forces. Our results brings up important implications for plasma fusion devices such as the tokamaks, the spheromaks and the reversed field pinch experiments, where the suppression of non-linearities is one of the most important objectives in the production of fusion energy. This work is also important to understand the self-organization of the solar wind but also the way MHD turbulent flows should be forced in numerical simulations.

The paper is structured as follows. All the necessary details on our DNS of driven MHD turbulent flows are provided in Sec. \ref{sec:dns}. Section \ref{sec:states} analyses the non-linear dynamics of the self-organized states. In particular, we focus on the dependence of the MHD flows on the forcing correlation time scale and its effect on the growth of helicities and the fate of the magnetic helical condensates. Individual statistics of the velocity and the magnetic field are presented in sec. \ref{sec:stats}. The diverse behaviour between the two fields dominates the statistics, where different dynamics are obeyed at different instances. Finally, in sec. \ref{sec:conclusion} we conclude by summarizing our findings and we discuss the implications of our work on the understanding of the self-organized processes that are observed in fusion plasma devices and in astrophysical observations.

\section{\label{sec:dns}Numerical methods}
Our study is based on numerical simulations of the MHD equations
\begin{align}
 (\pd_t - \nu \bm\Delta) \bm u &= (\bm u \times \bm \omega) + (\bm j \times \bm b) - \grad P + \bm f_u,
 \label{eq:ns} \\
 (\pd_t - \mu \bm\Delta) \bm b &=  \grad \times (\bm u \times \bm b) + \bm f_b, 
 \label{eq:induction}
\end{align}
where the vorticity $\bm \omega \equiv \grad \times \bm u$, the current density $\bm j \equiv \grad \times \bm b$, $P$ is the pressure, $\nu$ is the kinematic viscosity, $\mu$ is the magnetic diffusivity, $\bm f_u$ is the mechanical and $\bm f_b$ the electromagnetic external forces. Using the pseudo-spectral method, we numerically solve Eqs. \eqref{eq:ns} and \eqref{eq:induction} in a three-dimensional periodic box of size $2\pi$, satisfying $\grad \sdot \bm u = \grad \sdot \bm b = 0$. Aliasing errors are removed using the $2/3$ dealiasing rule, i.e. wavenumbers $k_{min}=1$ and $k_{max}=N/3$, where $N$ is the number of grid points in each Cartesian coordinate. For more details on the numerical code see \cite{mpicode05a,hybridcode11}.

In our simulations, the velocity and the magnetic field are forced at wavenumbers $k = 1$ and 2 with random phases. 
The random external forces $\bm f_u$ and $\bm f_b$ are normalised such that the forcing amplitude $|\bm f_u| = |\bm f_b| = |\bm f| = 1$ for all runs.
The forces have zero helicities, i.e. $\avg{\bm f_{u,b} \sdot \grad \times \bm f_{u,b}}_{_V} = \avg{\bm f_u \sdot \bm f_b }_{_V} = 0$ 
and their random phases change every time interval $\tau_c$ (where $\tau_c = \infty$ implies constant phases in time). The forcing correlation time scale $\tau_c$ is our parameter in this problem and is compared with $\tau_f \equiv (k_{min}|\bm f|)^{-1/2}$.
The energies of the initial conditions are chosen to be in equipartition (viz. $E_u = E_b = 0.5$) unlike in studies of relaxation processes 
\cite{tingetal86,striblingmatthaeus91}, where the initial conditions were chosen to have a tendency towards a particular relaxation state 
(i.e. force-free or Alfv\'enic). All the necessary parameters of our DNS are tabulated in Table \ref{tbl:dnsparam}. The magnetic Prandtl 
number is unity (i.e. $\nu = \mu$) for all the runs.

\begin{table}[!ht]
  \caption{Numerical parameters of the DNS. Note that $T_{tot}$ is the total run time.}
  \label{tbl:dnsparam}
   \begin{ruledtabular}
  \centering
    \begin{tabular}{*{4}{c}}
     \textbf{N} & $\bm \tau_c/\bm \tau_f$ & $\bm \nu = \bm \mu$ & $\bm T_{tot}/\bm \tau_f$ \\
     \hline
        64 & $\infty$ &  $1 \times 10^{-2}$ & 1550 \\
       128 &      0.5 &  $5 \times 10^{-3}$ & 1000 \\
       128 &      1.0 &  $5 \times 10^{-3}$ &  950 \\
       128 &      2.0 &  $5 \times 10^{-3}$ & 1000 \\
       128 &      4.0 &  $5 \times 10^{-3}$ & 1000 \\
       128 &      8.0 &  $5 \times 10^{-3}$ & 1000 \\
       128 & $\infty$ &  $5 \times 10^{-3}$ &  250 \\
       256 & $\infty$ &  $2 \times 10^{-3}$ &  105 \\
       512 & $\infty$ &  $9 \times 10^{-4}$ &   40 \\
    \end{tabular}
  \end{ruledtabular}
\end{table}

\section{\label{sec:states}Force-free, Alfv\'enic and Beltrami asymptotic states}

\begin{figure}[!ht]
  \includegraphics[width=8.5cm]{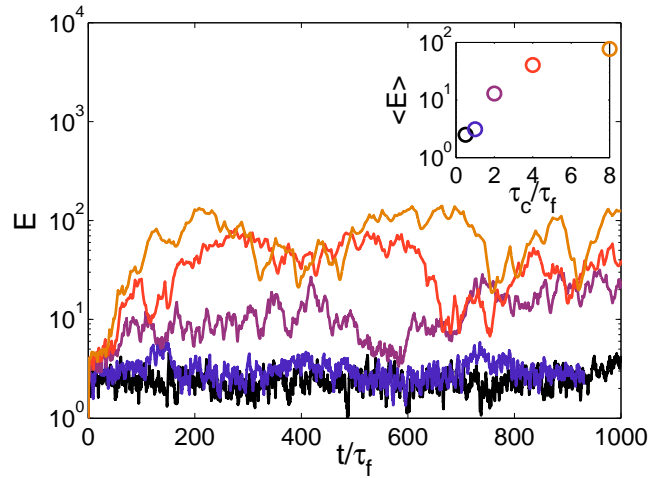}
  \caption{Time-series of total energy with different forcing correlation time scales. The inset shows the value of the time average total kinetic energy with respect to $\tau_c/\tau_f$.}
  \label{fig:timeseries}
 \end{figure}

To start with, we consider the temporal evolution of our flows with different forcing correlation time scales.
Remarkably, as $\tau_c/\tau_f$ increases we observe the amplitude of all space averaged quadratic quantities to increase substantially. This is depicted indicatively by the time-series of the total energy in Fig. \ref{fig:timeseries} plotted on a logarithmic scale. 
The inset shows the mean value of total energy $\avg{E}_t$ with respect to $\tau_c/\tau_f$.
Here, the angle brackets $\avg{\sdot}_t$ denote temporal averages.
As the forcing correlation time scale increases from $\tau_c/\tau_f = 0.5$ to 8, 
the mean value of total energy $\avg{E}_t$ increases by almost two orders of magnitude
and strongly fluctuates varying by an order of magnitude. Note that the time-series are characterized by dynamical time scales much greater than $\tau_c$, $\tau_f$ and the non-linear time scale $\tau_{_{NL}} \equiv 1/(k_{min}\avg{|\bm u|^2}_{_V}^{1/2})$. 
This behaviour indicates self-organisation in our MHD turbulent flows, which will become more obvious later on in our analysis. Due to these long dynamical time scales and the large variations we are compelled to integrate very far in time to obtain converged statistics and consequently deal with moderate resolutions (see Table \ref{tbl:dnsparam}).

\begin{figure}[!ht]
 \begin{subfigure}{4.25cm}
   \includegraphics[width=\textwidth]{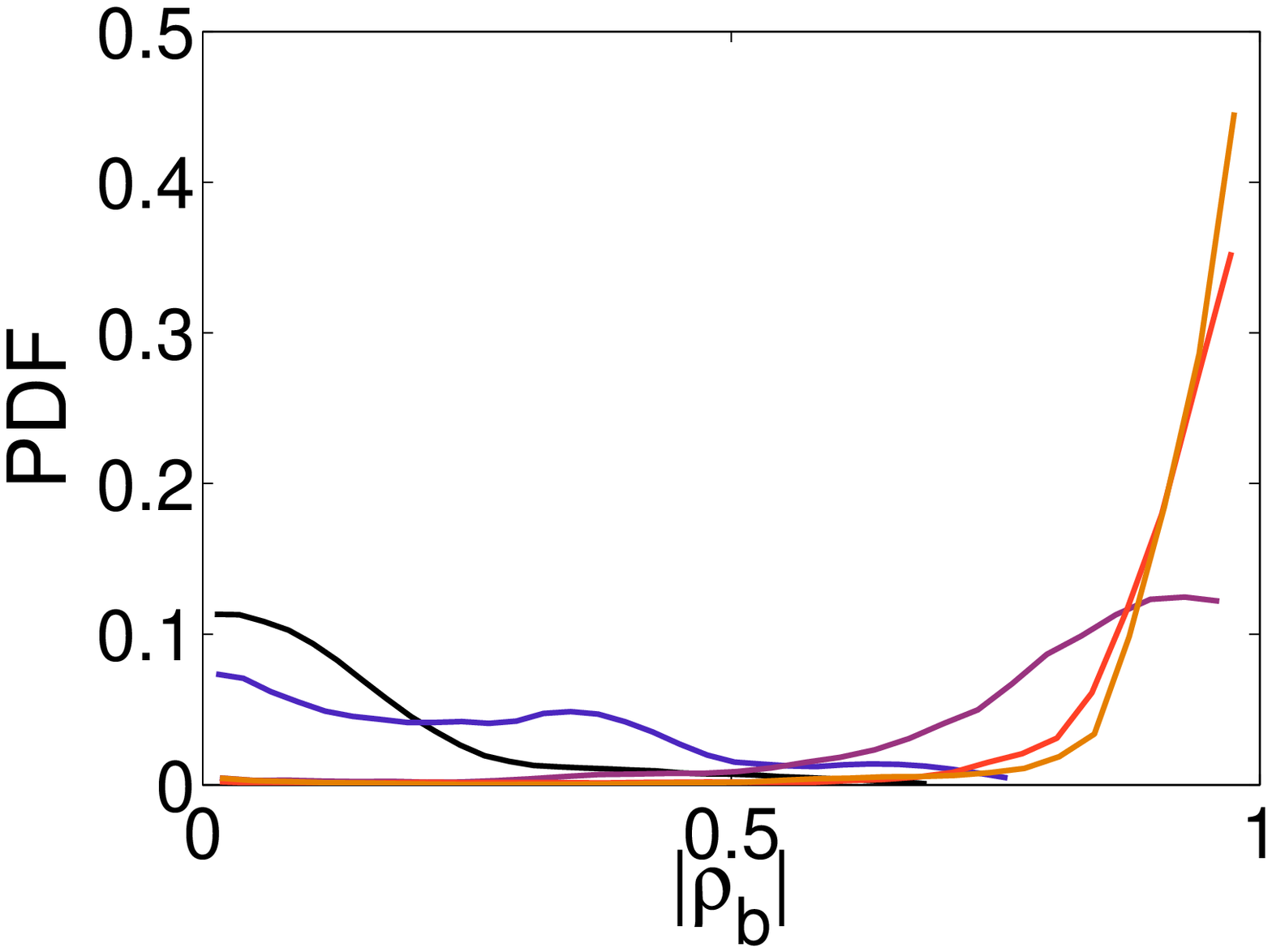}
   \caption{}
   \label{fig:pdfHm}
 \end{subfigure}
 \begin{subfigure}{4.25cm}
   \includegraphics[width=\textwidth]{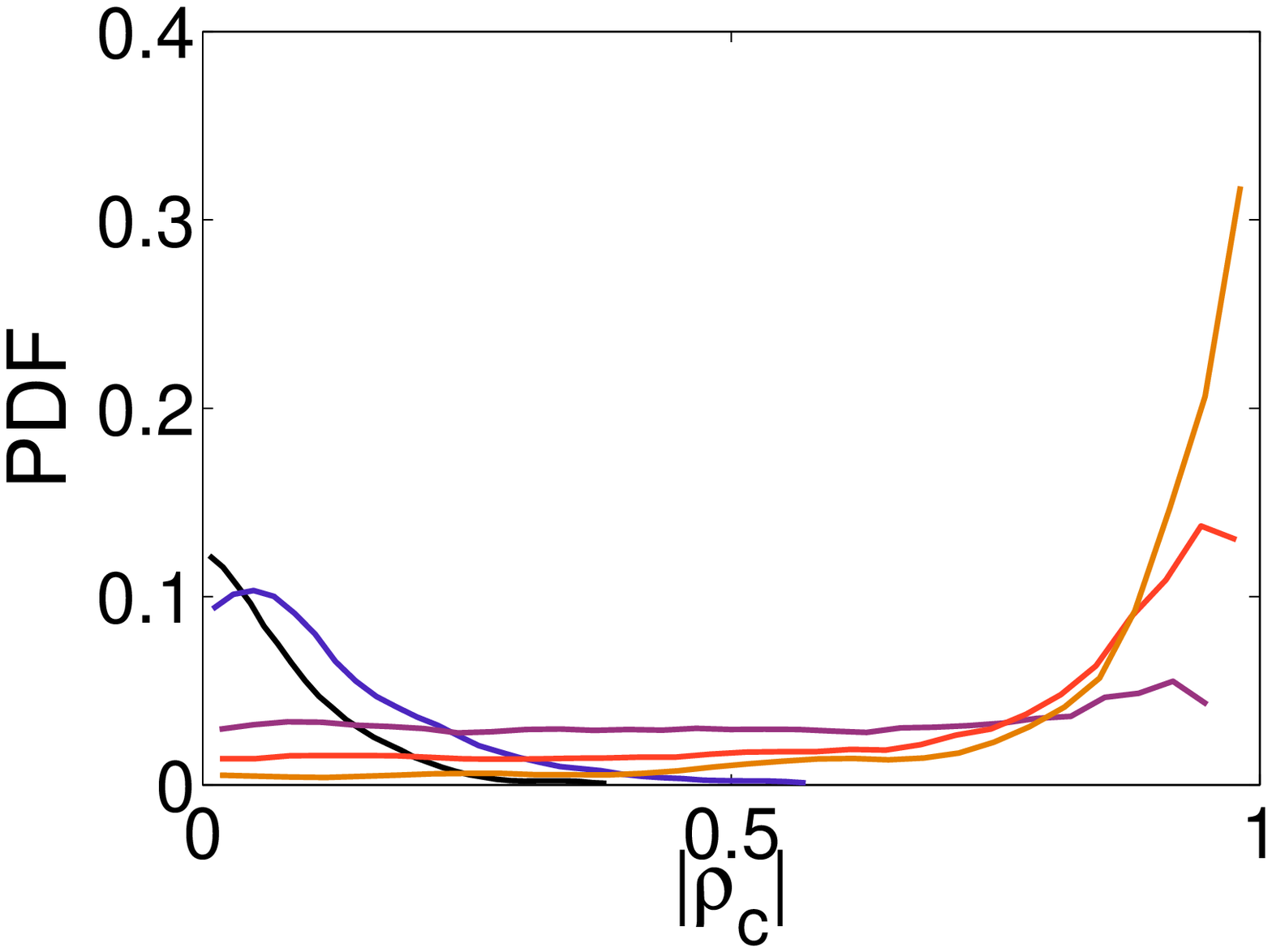}
   \caption{}
   \label{fig:pdfHc}
 \end{subfigure}
 \begin{subfigure}{4.25cm}
   \includegraphics[width=\textwidth]{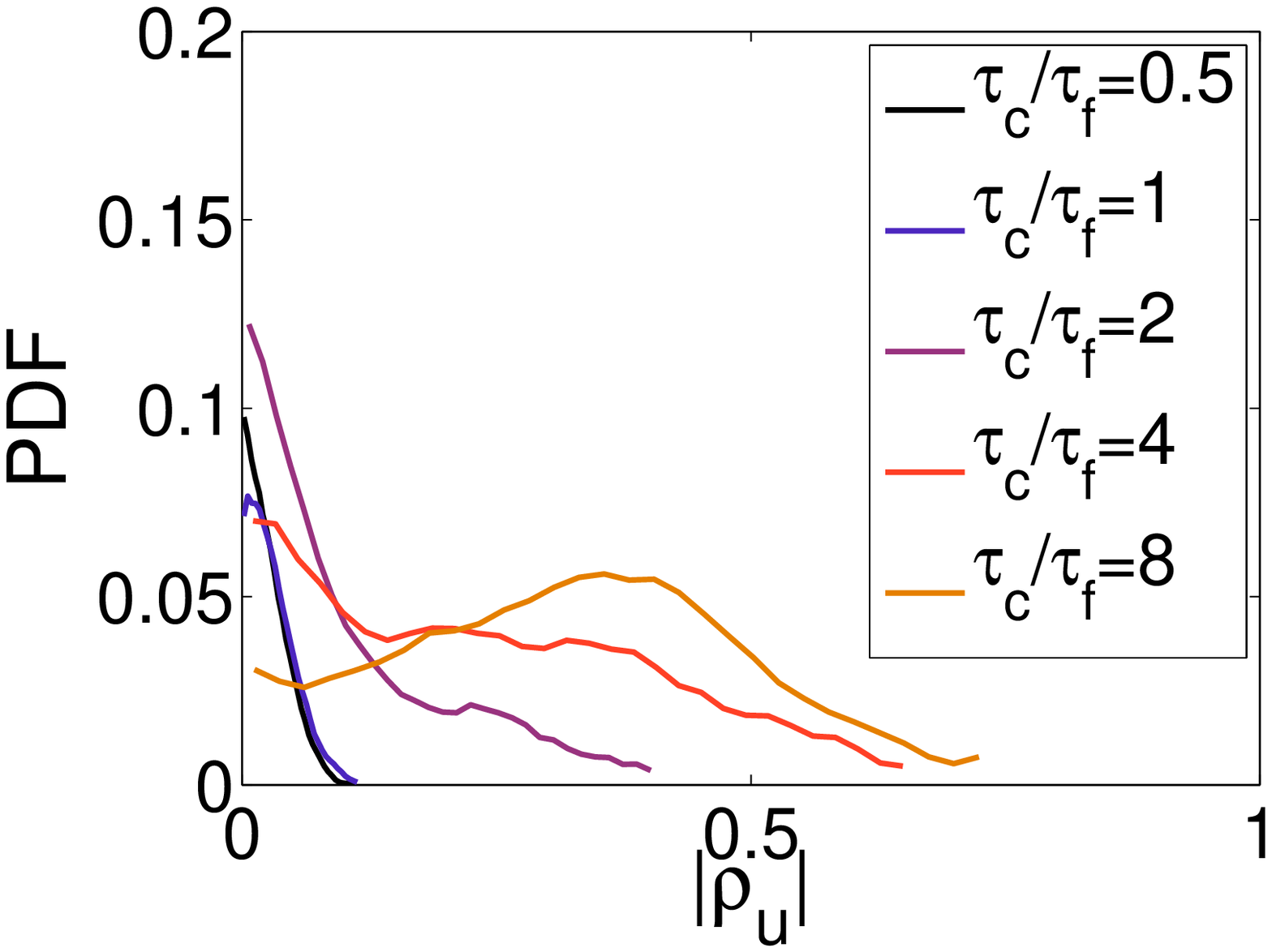}
   \caption{}
   \label{fig:pdfHk}
 \end{subfigure}
  \caption{Probability Density Functions of the absolute value of the normalised 
           (a) magnetic helicity  $|\rho_b|$, 
           (b) cross    helicity  $|\rho_c|$ and 
           (c) kinetic  helicity  $|\rho_u|$.}
  \label{fig:pdfs}
 \end{figure}

In Fig. \ref{fig:pdfs} we present the Probability Density Functions (PDFs) of the time series of the normalised cross, magnetic and kinetic helicities
\begin{align}
 \rho_c &\equiv H_c/(\avg{|\bm u|^2}_{_V}^{1/2}\avg{|\bm b|^2}_{_V}^{1/2}), \\
 \rho_b &\equiv k_{min}H_b/\avg{|\bm b|^2}_{_V}^{1/2}, \\
 \rho_u &\equiv H_u/(\avg{|\bm u|^2}_{_V}^{1/2}\avg{|\bm \omega|^2}_{_V}^{1/2}),
\end{align}
respectively. 
For low values of $\tau_c/\tau_f$ the PDFs of the normalized helicities are 
peaked around 0 as it is expected (see Fig. \ref{fig:pdfs}). 
However, for higher values of $\tau_c/\tau_f$ the PDFs becomes broader and shallower. For $\tau_c/\tau_f \geq 4$ the PDF of magnetic helicity in Fig. \ref{fig:pdfHm} peaks at 1 indicating the reach of a force-free state $\bm j \propto \bm b$ and thus $\bm j \times \bm b = 0$. The PDF of the normalized cross helicity in Fig. \ref{fig:pdfHc} also peaks at 1 for $\tau_c/\tau_f=8$ indicating
full Alfv\'enisation of the flow (i.e. $\bm u = \pm \bm b$). 
In the case of the normalized kinetic helicity, the PDFs with low $\tau_c/\tau_f$ values are peaked at 0 and they increase significantly for 
higher $\tau_c/\tau_f$ values without reaching full Beltramisation (i.e. $|\rho_u| = 1$ or $\bm u = \pm \bm \omega$) even for $\tau_c/\tau_f = 8$ (see Fig. \ref{fig:pdfHk}).
The preference for these highly helical states, where non-linearities (and the cascade to small scales) are quenched, are related to the excess of energy that was observed in Fig. \ref{fig:timeseries}.

\begin{figure}[!ht]
 \begin{subfigure}{4.25cm}
   \includegraphics[width=\textwidth]{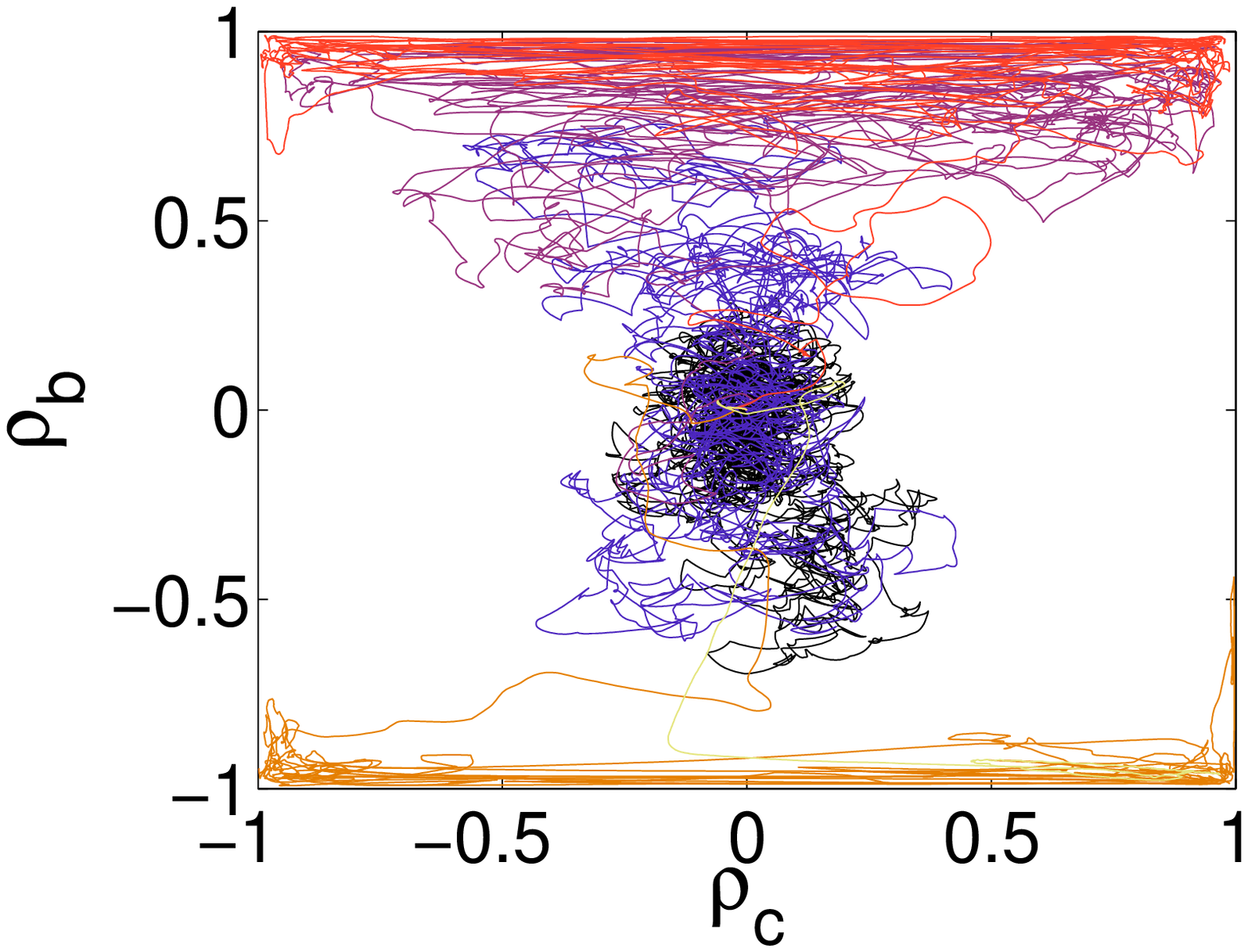}
   \caption{}
   \label{fig:HmHc}
 \end{subfigure}
 \begin{subfigure}{4.25cm}
   \includegraphics[width=\textwidth]{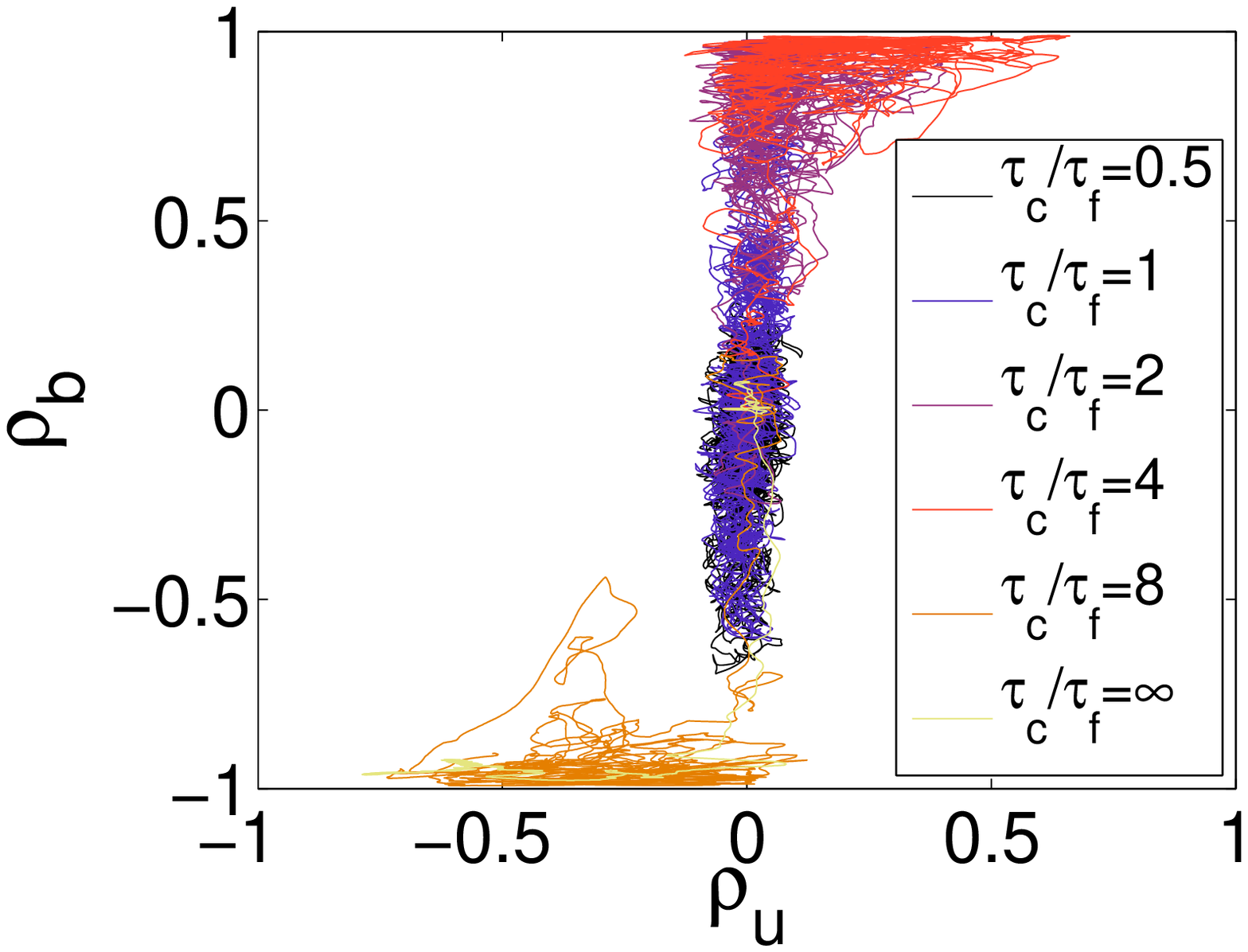}
   \caption{}
   \label{fig:HmHk}
 \end{subfigure}
 \begin{subfigure}{4.25cm}
   \includegraphics[width=\textwidth]{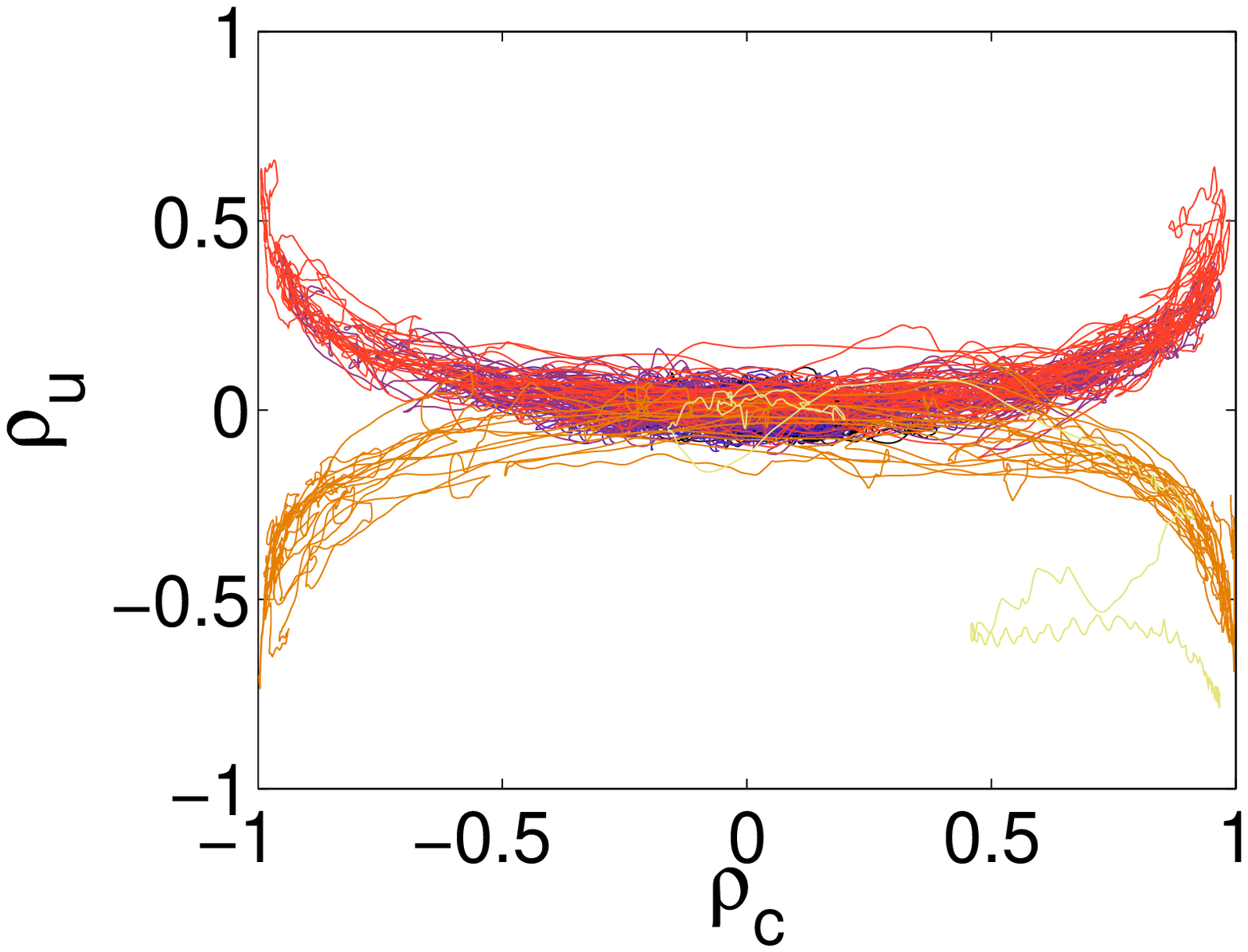}
   \caption{}
   \label{fig:HkHc}
 \end{subfigure}
 \begin{subfigure}{4.25cm}
   \includegraphics[width=\textwidth]{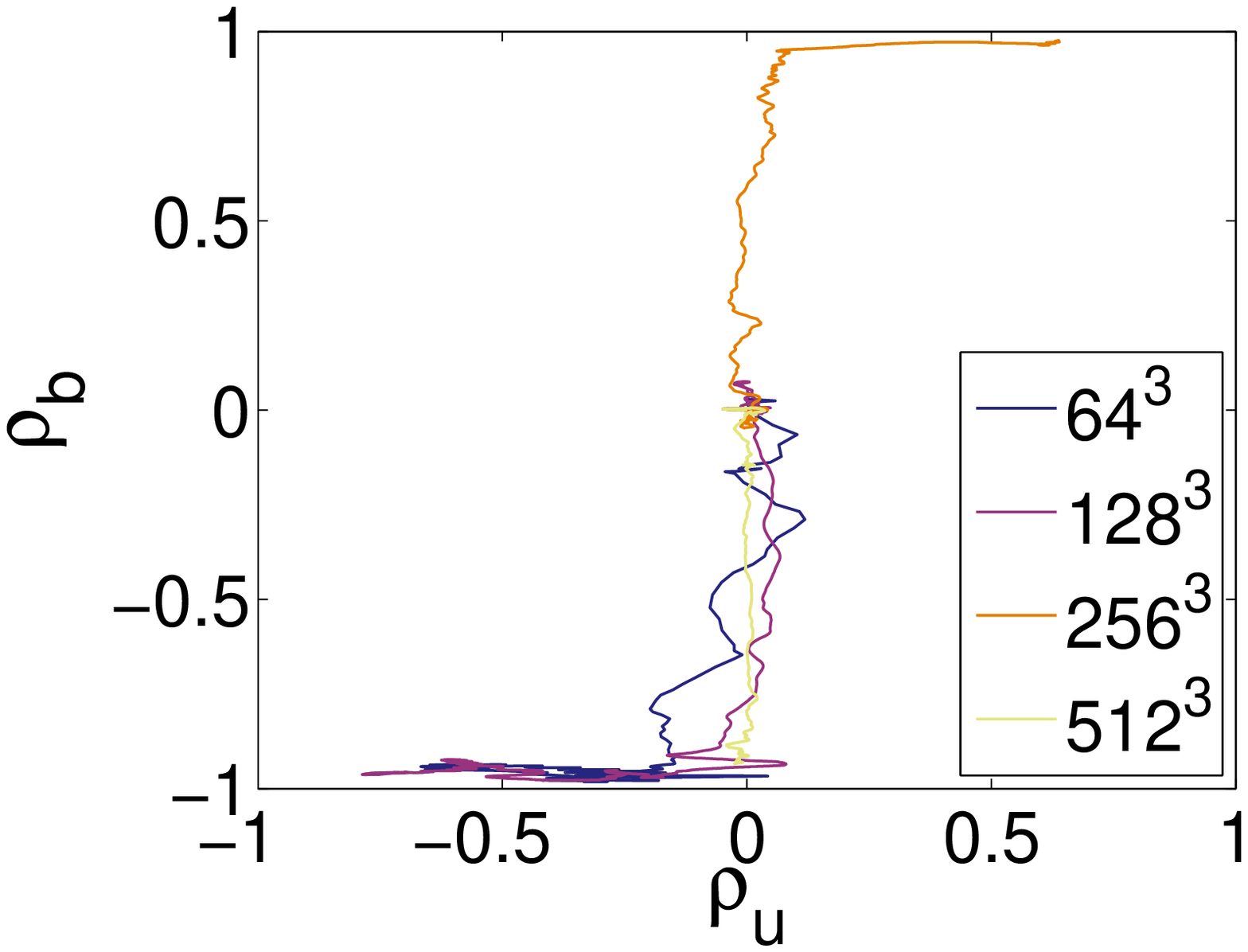}
   \caption{}
   \label{fig:HmHcinf}
 \end{subfigure}
  \caption{Phase sub-space of (a) $\rho_b$ and $\rho_c$, (b) $\rho_b$ and $\rho_u$, (c) $\rho_u$ and $\rho_c$ for flows with different $\tau_c/\tau_f$. (d) Phase sup-space of $\rho_b$ and $\rho_u$ for flows with different resolutions and $\tau_c/\tau_f = \infty$.}
  \label{fig:attractors}
 \end{figure}

Now we examine the evolution of our flows in a three-dimensional phase space composed by the three normalized helicities. Figure \ref{fig:attractors} demonstrates the time evolution of the solutions of our flows and their dependence on the forcing correlation time scale $\tau_c/\tau_f$,
which varies from 0.5 to $\infty$ (see Table \ref{tbl:dnsparam}). In particular, Fig. \ref{fig:HmHc} shows the phase sub-space of the normalized magnetic and cross 
helicities, where we observe that the flows with $\tau_c/\tau_f = 0.5$ and 1 mainly oscillate around the 0 origin with some excursions away from 0.
As the forcing correlation time scale increases from $\tau_c/\tau_f = 2$ to $\infty$, we notice that these excursions are trapped into two attracting solutions 
$\rho_b \simeq \pm 1$, oscillating between $-1 \leq \rho_c \leq 1$ for long times (i.e. $T_{tot}/\tau_f \simeq 1000$). In other words, the solution is trapped 
into force-free states (i.e. $\bm j \propto \bm b$) while it oscillates between the two Alfv\'enic states (i.e. $\bm u = \pm \bm b$). 
For the runs with long correlation times $\tau_c$,  we do not observe noise driven escape from one basin of attraction to the other.
We expect the probability to exit from one metastable potential well to the other 
to be vanishingly small due to the fact that the noise in the system is not enough to escape
the solution from either basin of attraction due to the suppression of the non-linearities. On the other hand, we observe noise driven
escape of the system between the two attractors of the Alfv\'enic states (i.e. $\rho_c = \pm 1$). It is of interest to mention here that the choice between the two asymptotic fully magnetically helical states depends sensitively on the initial conditions. A small 
perturbation in magnetic helicity can lead the system to be trapped either in the $\rho_b=1$ or $\rho_b=-1$ state, much like a small variation in the initial conditions of a coin-toss experiment can alter the final results from head to tails \cite{keller86}.

The phase sub-space of the normalized magnetic and kinetic helicity (see Fig. \ref{fig:HmHk}) 
reveals two states with positive and negative $H_u$ for high enough values of $\tau_c/\tau_f$. 
Note that kinetic helicity grows only when $|\rho_b| > 0.5$ and always has the same sign with magnetic helicity. This implies that as the magnetic field becomes strong it forces the flow to become Alfv\'enized (i.e. $\bm u \propto \bm b$) fluctuating between positive and negative $H_c$ but always keeping the same sign for $H_u$ and $H_b$ (see Fig. \ref{fig:attractors}) because kinetic and magnetic helicities are invariant under $\bm u \to -\bm u$ and $\bm b \to -\bm b$ transformations, respectively. For completeness of the three-dimensional phase space of normalized helicities, the phase sub-space of kinetic and cross-helicity is presented in Fig. \ref{fig:HkHc}. Here, the two attractors of positive and negative $H_u$ are also evident, with the flows of high forcing correlation time scales to oscillate for long times between the $\bm u = \pm \bm b$ solutions, whereas the flows with low $\tau_c/\tau_f$ to oscillate around $\rho_u = 0$.


The same conclusions are more clearly depicted in Fig. \ref{fig:HmHcinf} where $\rho_u$ and $\rho_b$ is shown for all the resolutions considered in this study (i.e. $64^3$ to $512^3$) with $\tau_c/\tau_f = \infty$. Therefore, we expect this behaviour to persist for any Reynolds number flow with high enough time-correlated external forces. Note that the particular case of fully time-correlated external forces (i.e. $\tau_c/\tau_f = \infty$) reaches higher values of normalized kinetic helicity than the rest of the flows. In this case, one could hypothesize that if this flow will be integrated very far in time it will reach a fully Beltrami asymptotic state (i.e. $\bm u = \pm \bm \omega$). However, even for the $64^3$ run, which was integrated for longer time scales (i.e. $T_{tot}/\tau_f \simeq 1550$), the flow did not reach full Beltramisation. So, it is not clear if a fully kinetic helical state will ever be reached for $\tau_c/\tau_f = \infty$ and $T_{tot}/\tau_f \gg 1$.

\begin{figure}[!ht]
  \includegraphics[width=8.5cm]{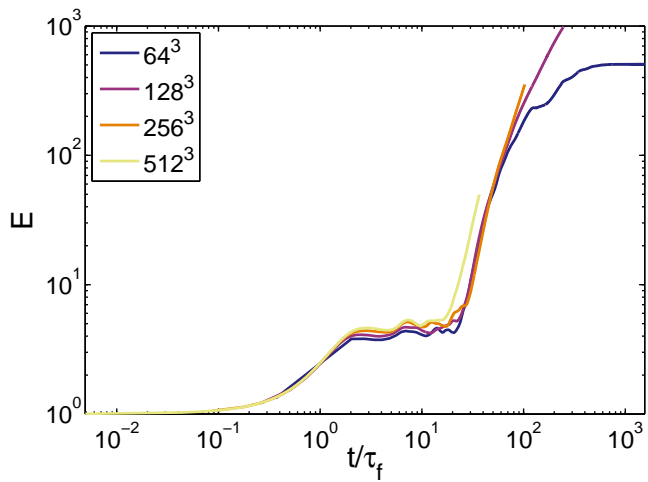}
  \caption{Time-series of the total energy for flows with different resolutions and fully time-correlated external forces, i.e. $\tau_c/\tau_f = \infty$.}
  \label{fig:et}
 \end{figure}

At this point, we should identify the fate of these helical condensates at $T_{tot}/\tau_f \gg 1$ for fully time-correlated forces in our simulations of driven MHD turbulence. This is clearly illustrated by plotting the time series of the total energy $E$ for the different resolutions that we have considered with $\tau_c/\tau_f = \infty$ (see Fig. \ref{fig:et}).
For early times, all the runs seem to asymptote towards a steady state solution (i.e. $t/\tau_f \simeq 10$). However, when we integrate further in time the solutions deviate abruptly
from this temporary steady state towards their fully helical states, which eventually asymptote to the laminar attractor, i.e. the flows reach their viscous limit. Note that the reach of the viscous asymptotic limit becomes prohibitive as resolution increases, because the time-step becomes vanishingly small, based on a CFL criterion, in order to capture the required dynamics.

The question that now arises is why these helical states manifest. This manifestation seems to be intimately connected with the inverse cascade of magnetic helicity \cite{frischetal75}. The presence of the electromagnetic force $\bm f_b$ induces the condensation of $E_b$ into a large scale helical magnetic field (magnetic condensate), which is a consequence of the inverse cascade of $H_b$. These magnetic condensates can be shown to be unconditionally stable for any Reynolds number (see section \ref{sec:appendix}) in contrast to Beltrami states in 3D helical hydrodynamic turbulence, which are unstable and energy is cascaded forward. 
%
So, for $\tau_c/\tau_f \rightarrow \infty$ the flow will be attracted to the laminar stable states for any Reynolds number as Fig. \ref{fig:et} also indicates. For finite but large enough $\tau_c/\tau_f$ these states will be also approached, while for small $\tau_c/\tau_f$ the external forces change before the system has time to reach a self-organised state.
We remark that self-organisation is induced in MHD turbulence as soon as an electromagnetic force $\bm f_b$ is involved in the equations, with either $|\bm f_b| \sim |\bm f_u|$ or $|\bm f_b| \gg |\bm f_u|$. Otherwise, when $\bm f_b = 0$ (i.e. dynamo) or $|\bm f_b| \ll |\bm f_u|$, helicities do not grow even for fully time-correlated forces, reaching a statistically steady state with helicities' fluctuations around zero, as it is expected.


%
\section{\label{sec:stats}Statistics of self-organised states}
\begin{figure}[!ht]
 \begin{subfigure}{8.5cm}
   \includegraphics[width=\textwidth]{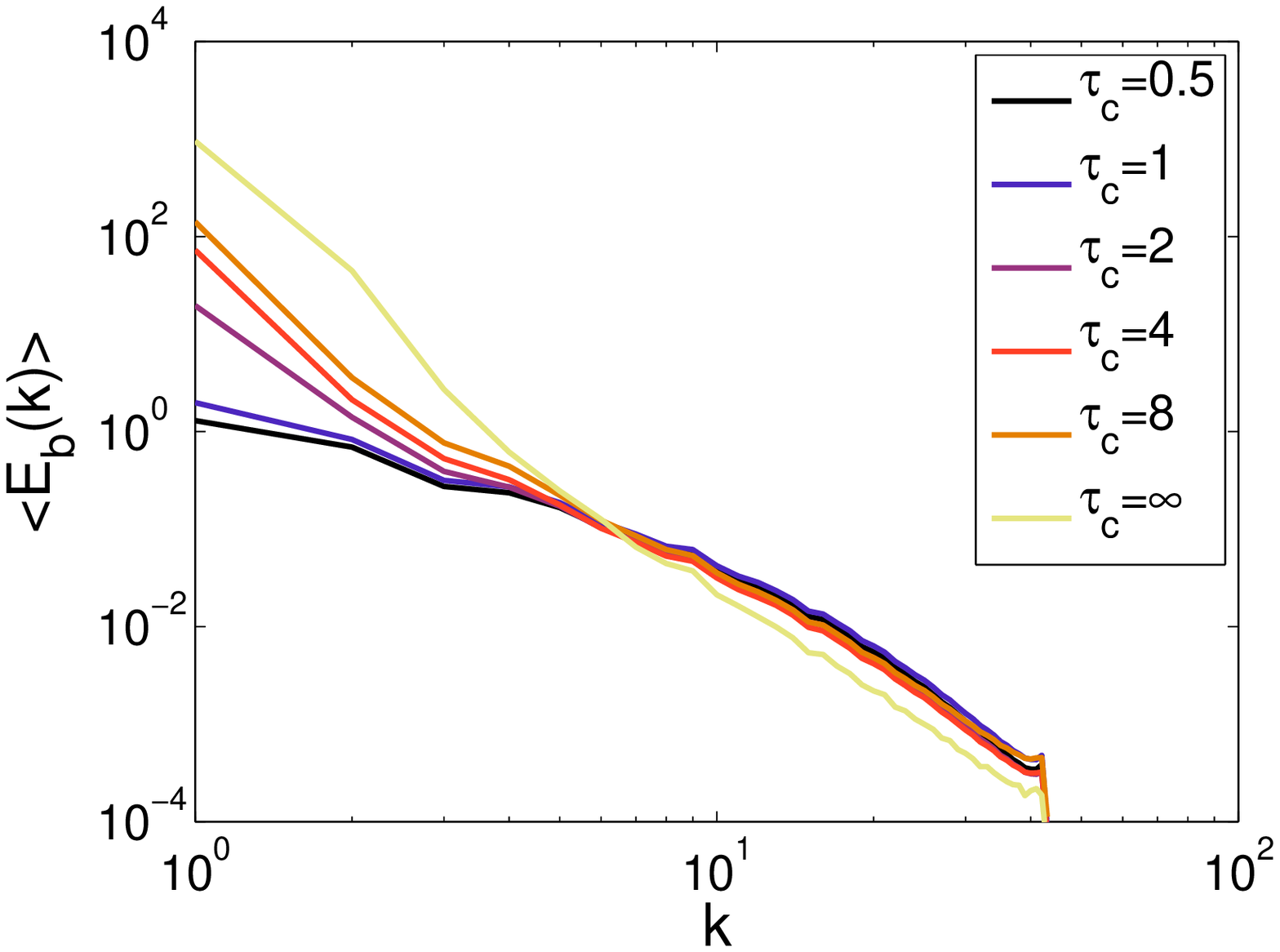}
   \caption{}
   \label{fig:Eb}
 \end{subfigure}
 \begin{subfigure}{8.5cm}
   \includegraphics[width=\textwidth]{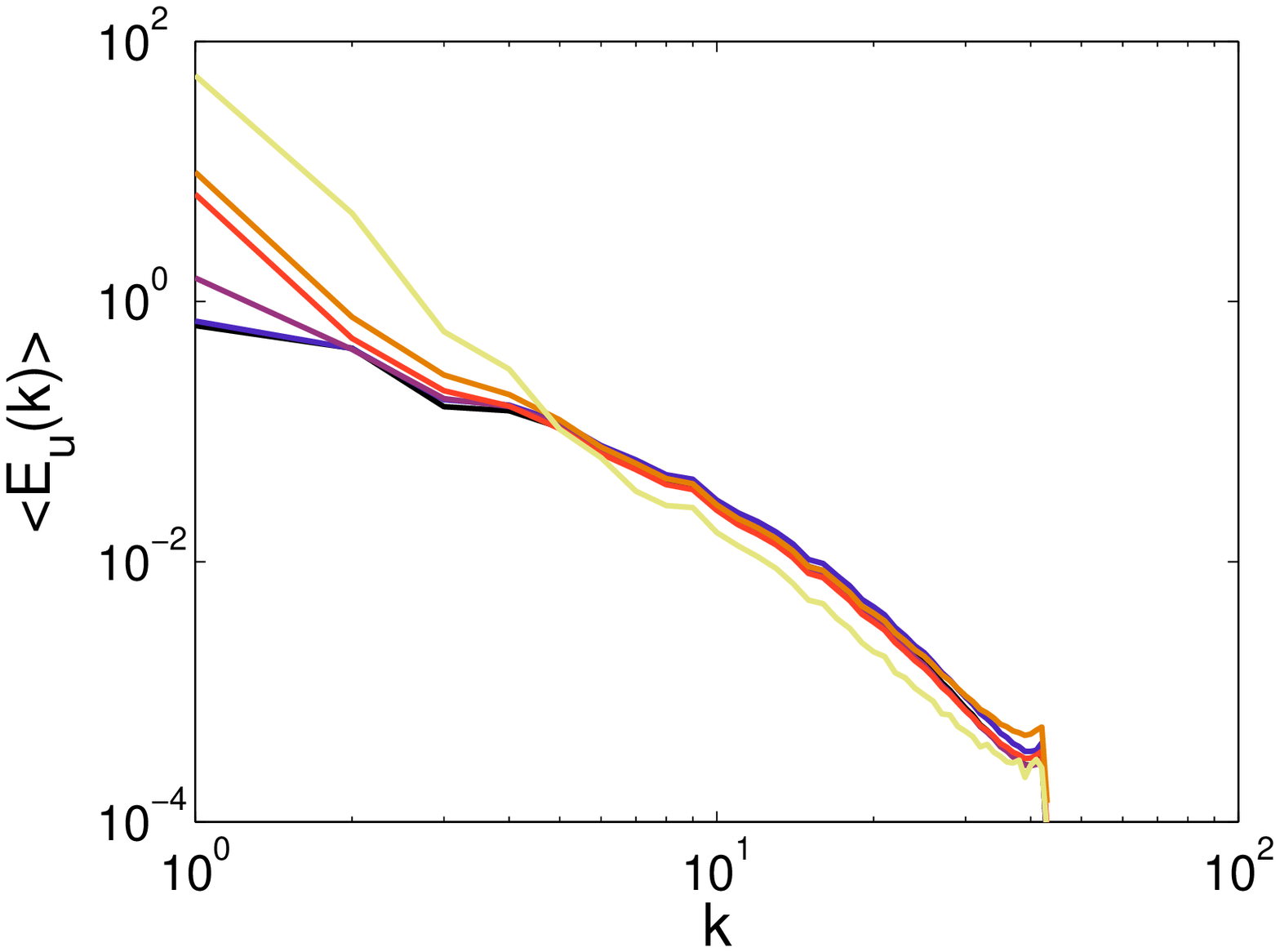}
   \caption{}
   \label{fig:Eu}
 \end{subfigure}
  \caption{Time averaged (a) magnetic and (b) kinetic energy spectra for different values of $\tau_c/\tau_f$.}
  \label{fig:spectra}
 \end{figure}

In contrast to the relaxation processes in selective decay, we are able to have statistics of the asymptotic force-free, Alfv\'enic and Beltrami states. Figure \ref{fig:spectra} presents the magnetic and kinetic energy spectra averaged in time for the flows with different $\tau_c/\tau_f$.
It is obvious that both the magnetic and kinetic energy increases significantly at low wavenumbers as $\tau_c/\tau_f$ increases, 
implying the creation of helical condensates at large scales.
Comparing the magnetic and the kinetic energy spectra we observe that the magnetic helical condensates dominate the large scales 
of the flow with $\avg{E_b}_t > \avg{E_u}_t$ at low wavenumbers. As the external forces become more correlated in time, the ratio 
$\avg{E_b}_t / \avg{E_u}_t$ presents also a considerable increase at large scales. 
The slope of the energy spectrum appears to be very steep, particularly for high $\tau_c/\tau_f$ values, and much steeper than typical turbulence theory exponents $-5/3$ and $-3/2$. Note that large fluctuations on the spectral exponent were observed throughout the run, with steep exponents during the condensate phases and more turbulent-like exponents within intermediate times. Therefore, the presence of the helicity condensates can clearly affect the time averaged energy spectrum even when their duration is short (e.g. $\tau_c/\tau_f = 2$ in our flows). Due to the relatively small resolution of these simulations the statements about the spectral exponents are only qualitative.

\begin{figure}[!ht]
 \begin{subfigure}{8.5cm}
   \includegraphics[width=\textwidth]{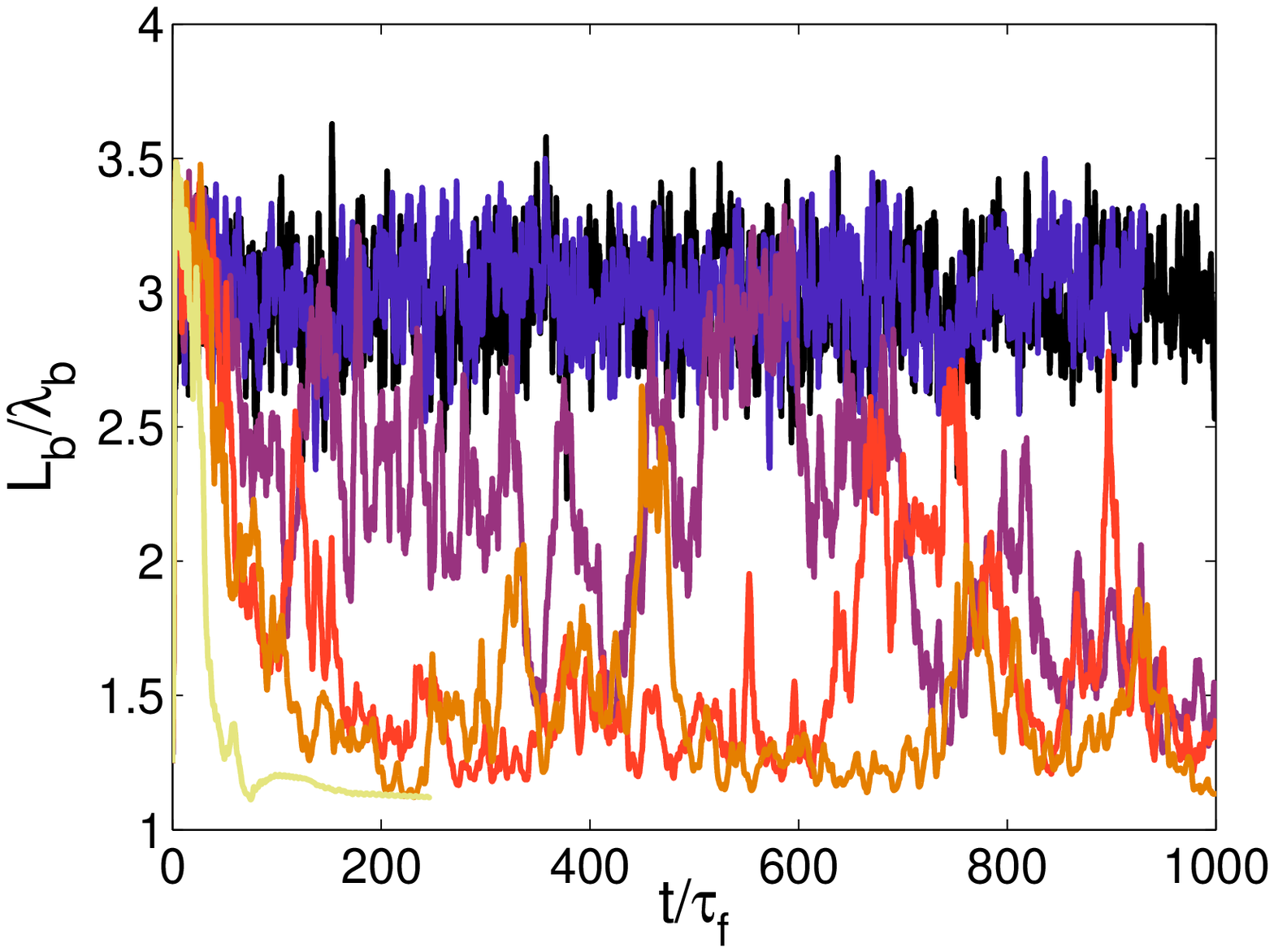}
   \caption{}
   \label{fig:Lb}
 \end{subfigure}
 \begin{subfigure}{8.5cm}
   \includegraphics[width=\textwidth]{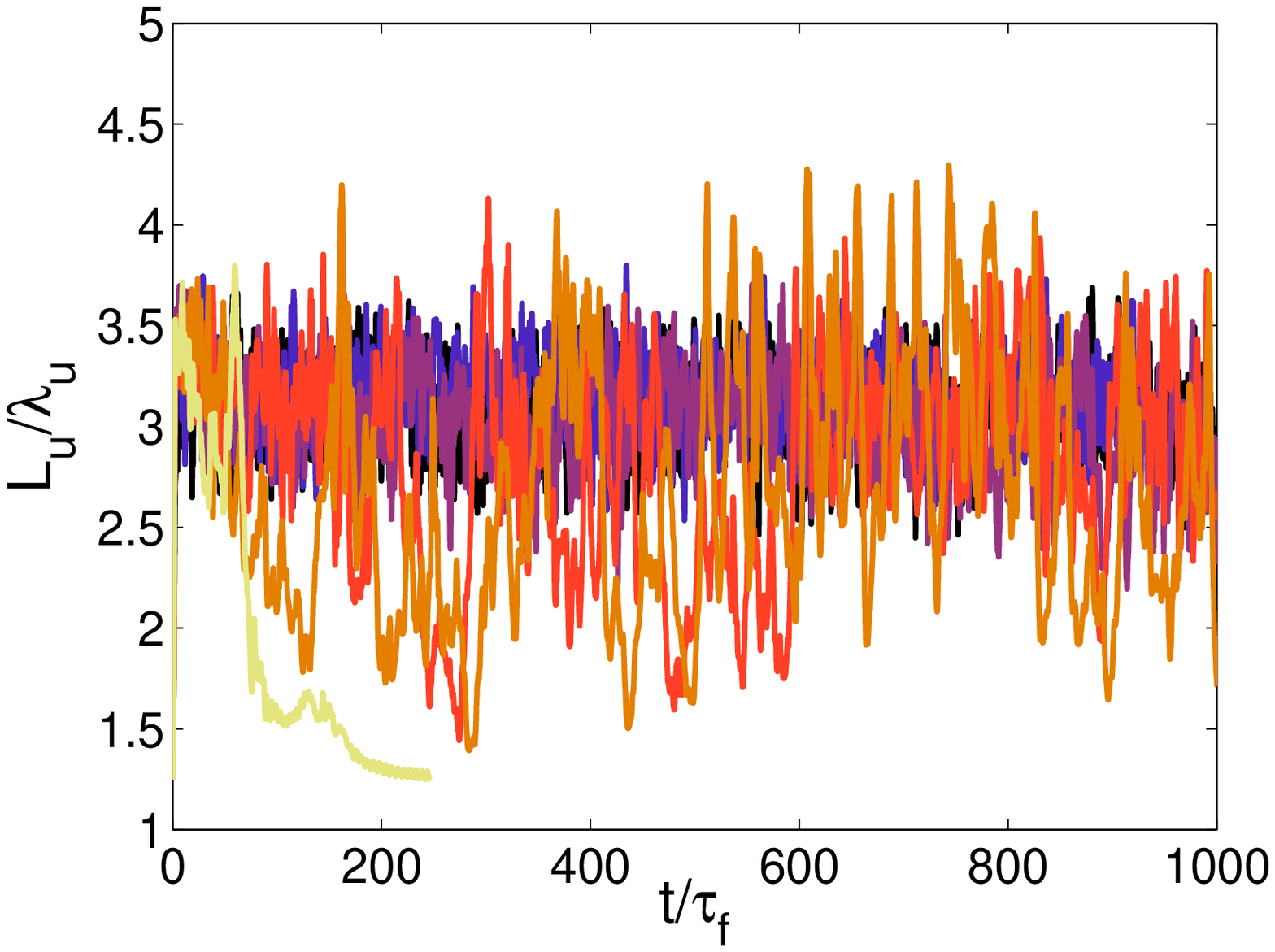}
   \caption{}
   \label{fig:Lu}
 \end{subfigure}
  \caption{Time-series of the ratio of the integral length scale to the Taylor micro-scale of (a) the magnetic field and (b) the velocity field for different forcing correlation time scales from $\tau_c/\tau_f = 0.5$ to $\infty$. The curves follow the same labelling with Fig. \ref{fig:spectra}.}
  \label{fig:scales}
 \end{figure}

Alternatively, to show the existence of the large fluctuations that could affect the energy spectrum we analyse the dynamics of the large and small scales. So, we compute the corresponding integral scales of the velocity and the magnetic field individually
\begin{equation}
 L_{u,b} \equiv \frac{3\pi}{4} \frac{\int k^{-1}E_{u,b}(k)dk}{\int E_{u,b}(k)dk}
\end{equation}
and Taylor micro-scales
\begin{equation}
 \lambda_{u,b} \equiv \lrbig{\frac{5 \int E_{u,b}(k)dk}{\int k^2E_{u,b}(k)dk}}^{1/2}. 
\end{equation}
In Fig. \ref{fig:scales} we illustrate the time-series of the ratios of the integral length scale to Taylor micro-scale of the magnetic field and the velocity field. 
When energy is concentrated on the largest scale of the system $k = 1$ the ratio $L_{u,b}/\lambda_{u,b} = \frac{3\pi}{4\sqrt{5}} \simeq 1.05$ and thus the flow exhibits laminar behaviour while when $L_{u,b}/\lambda_{u,b} \gg 1$ a turbulent scaling is expected and dissipation is dominated by small scales.
The difference in the dynamics of the velocity and the magnetic field is striking as the forcing correlation time scale varies. 
On one hand, the magnetic field reaches different asymptotic states for different values of $\tau_c/\tau_f$ (see Fig. \ref{fig:Lb}) and it moves toward the laminar attractor, i.e. $L_b/\lambda_b \rightarrow 1$, as $\tau_c/\tau_f$ becomes high enough. The time-series of $L_b/\lambda_b$ is characterized by very rare and extreme events away from the laminar attractor for high values of $\tau_c/\tau_f$. 
On the other hand, even though the amplitude of the fluctuations of $L_u/\lambda_u$ increases significantly with respect to the increase of the forcing correlation time scale, the mean value of $L_u/\lambda_u$ remains almost at the same level except for the run with $\tau_c/\tau_f = \infty$ (see Fig. \ref{fig:Lu}).
This diverse behaviour between the velocity and the magnetic field dominates the statistics and this was also depicted in the time-averaged energy spectra in Fig. \ref{fig:spectra}. The large amplitude and rare fluctuations in the time-series of $L_b/\lambda_b$ and $L_u/\lambda_u$ imply that different spectra are obtained in different instances with this becoming more pronounced for the high values of the forcing correlation time scale. This behaviour is expected for any Reynolds number for high enough time-correlated external forces.

\section{\label{sec:conclusion}Discussion and Conclusions}
In this work, we consider MHD turbulent flows driven by kinetic and electromagnetic external forces. By increasing the time-correlation of the external forces 
we demonstrate that MHD turbulent flows create large-scale condensates out of a sea of homogeneous turbulence. This self-organization is connected with the 
presence of an electromagnetic external force, which induces the condensation of magnetic energy into the large scale helical coherent structures as a consequence 
of the inverse cascade of magnetic helicity even though the forcing itself is not helical.
The magnetic energy at the large scales of the flow is always greater than the kinetic energy and the ratio of magnetic 
to kinetic energy grows as the forcing correlation time scale increases.

From our phase space analysis, we conclude that helicities increase as the forcing correlation time scale increases and drive the MHD flows toward a laminar attractor at large integration times. Flows with high enough value of $\tau_c/\tau_f$ reach different asymptotic states of 
different levels of helicities. The evolutions of these flows are governed by the force-free, Alfv\'enic and Beltrami states (i.e. $\bm u \propto \bm b \propto \bm \omega \propto \bm j$), which appear to be attracting solutions of the MHD equations when $\tau_c/\tau_f \rightarrow \infty$ for any 
Reynolds number. The existence of these attractors is intimately connected to the unconditional stability of the magnetic helical condensates for any Reynolds number. So, a predictive theory for self-organisation in electromagnetically driven MHD turbulent flows is plausible based on the fact that asymptotic laminar attractors govern the dynamics of such flows with long enough forcing correlation time scales.

Moreover, the dynamics of the velocity and the magnetic field present very diverse dynamics, which dominate the statistics. In other words, at different instances in time rare events provide very disparate energy spectra during the evolution of the MHD flows for high enough correlation time scales of the external forces. These results raise the issue of how a DNS of MHD turbulent flows should be forced in order to avoid any lack of universal results to be caused by the 
self-organisation of the flow. In particular the current study excludes the use of time-independent magnetic forcing for the study of steady state MHD turbulence in periodic boxes as in this case the flow will be attracted to the self-organised states.
%

Our study has perhaps important practical implications for many electromagnetically forced devices such as electromagnetic pumps in nuclear stations, fusion plasma 
devices such as the tokamaks, spheromaks and reversed field pinch experiments. Note that the self-organized states can persist for all time, if permitted by boundary 
conditions, as they lead to the cancellation of the non-linearities and the system toward the laminar attractor. The presence of coherent structures inside 
turbulence greatly affect the plasma diffusion process. Therefore, further study of the self-organization processes is vital in order to control the transport processes in fusion plasmas and hence to improve the plasma confinement. Moreover, an increased knowledge of these phenomena in MHD turbulence can shed light on the understanding of some basic phenomena in solar corona, such as flares and coronal heating, which occur in a region where observations cannot be performed.


%
\section{\label{sec:appendix}Appendix}
Here we show that magnetic helical condensates at the largest scale of MHD flows are unconditionally stable for any Reynolds number. We consider an arbitrary velocity field $\bm u'$ and a magnetic field with a large scale fully helical component $\bm B_0$ and a small scale component $\bm b'$, i.e. $\bm b = \bm B_0 + \bm b'$. The large scale magnetic field is such that
\begin{equation}
\grad \times \bm B_0 = \pm k_0 \bm B_0=\bm J_0,
\end{equation}
where $k_0$ is the smallest wavenumber in the domain. The small scale component $\bm b'$ is composed of all the Fourier modes such that $|\bm k| \ne k_0$ (i.e. all the Fourier modes except the ones for which belong to $B_0$).

Then, the Eqs. \eqref{eq:ns} and \eqref{eq:induction} for the small scale fields become
\begin{align}
 \pd_t \bm u' &= (\bm J_0 \times \bm b') + (\bm j' \times \bm B_0) + (\bm j' \times \bm b') + \nu \bm\Delta \bm u' \nonumber \\
              &+ (\bm u' \times \bm \omega') - \grad P
 \label{eq:ns2}
\end{align}
and
\begin{align}
 \pd_t \bm b' &=\grad \times (\bm u' \times \bm b') + \grad \times (\bm u' \times \bm B_0) - \mu \grad \times \bm j'.
 \label{eq:induction2}
\end{align}
To derive the equations for the averaged kinetic and magnetic energy, we multiply Eq. \eqref{eq:ns2} by $\bm u'$ and Eq. \eqref{eq:induction2} by $\bm b'$ and we integrate over the volume $V$ to obtain
\begin{align}
 \pd_t \avg{|\bm u'|^2} &= \avg{\bm u' (\bm J_0 \times \bm b')} + \avg{\bm u' (\bm j' \times \bm B_0)} + \avg{\bm u' (\bm j' \times \bm b')} \nonumber \\ 
                &+ \nu \avg{|\grad \bm u'|^2}
 \label{eq:eu}
\end{align}
and
\begin{align}
 \pd_t \avg{|\bm b'|^2} &= \avg{\bm j' (\bm u' \times \bm b')} + \avg{j' (\bm u' \times \bm B_0)} - \mu \avg{|\bm j'|^2}.
 \label{eq:eb}
\end{align}
where $\avg{.} \equiv \int_V \dd^3 x$.
By summing Eqs. \eqref{eq:eu} and \eqref{eq:eb}, we obtain
\begin{align}
 \pd_t (\avg{|\bm u'|^2} + \avg{|\bm b'|^2}) &= \avg{\bm u' (\bm J_0 \times \bm b')} - \nu \avg{|\grad \bm u'|^2} - \mu \avg{|\bm j'|^2} \nonumber \\
         &= k_0\avg{\bm u' (\bm B_0 \times \bm b')} - \nu \avg{|\grad \bm u'|^2} - \mu \avg{|\bm j'|^2}.
 \label{eq:et}
\end{align}

Let $\bm a \equiv \bm A_0 + \bm a'$ (where $\grad \times \bm A_0 = \bm B_0$ and $\grad \times \bm a' = \bm b'$),
be the vector potential.
Multiplying Eq. \eqref{eq:induction2} with $\bm a'$  and integrate over the volume $V$ we obtain
\begin{align}
 \pd_t \avg{\bm a' \sdot \bm b'} &=  \avg{\bm b' (\bm u' \times \bm B_0)} - \mu \avg{\bm j' \sdot \bm b'}.
 \label{eq:hb}
\end{align}
where $\avg{\bm a' \sdot \bm b'}$ is the magnetic helicity of the perturbation field.
Multiplying Eq. \eqref{eq:hb} by $k_0$ and then subtracting from Eq. \eqref{eq:et} leads to
\begin{equation}
\pd_t \mathcal{M} = - \nu \avg{|\grad \bm u'|^2} - \mu \avg{|\bm j'|^2} + k_0 \mu \avg{\bm j' \sdot \bm b'}
\label{eq:M}
\end{equation} 
where $\mathcal{M} \equiv \avg{|\bm u'|^2} + \avg{|\bm b'|^2} - k_0\avg{\bm a' \sdot \bm b'}$.
We now show that $\mathcal{M}$ is a non-negative functional as follows
\begin{align}
\mathcal{M} &=   \avg{|\bm u'|^2} + \avg{|\bm b'|^2} - k_0 \avg{\bm a' \sdot \bm b'} \\
             &\ge \avg{|\bm b'|^2} - k_0\avg{\bm a' \sdot \bm b'} \\
             &\ge \avg{|\bm b'|^2} - k_0\avg{|\bm b'|^2}^{1/2} \avg{|\bm a'|^2}^{1/2} \\
             &\ge 0 
\end{align}
where the last inequality comes from noting that 
\begin{align}
 \avg{|\bm a'|^2} &=   \sum_{|\bm k|>k_0} k^{-2}|\hat{\bm b}'_k|^2 \\
              &\le \sum_{|\bm k|>k_0} k_0^{-2} |\hat{\bm b}'_k|^2 \\
              &=    k_0^{-2}\avg{|\bm b'|^2}.
\end{align}
The equal sign holds only for $\bm u' = \bm b' = 0$. Following the same steps we can show that
\begin{equation}
 -\nu \avg{|\grad \bm u'|^2} -\mu \avg{|\bm j'|^2} + k_0 \avg{\bm j' \sdot \bm b'} \le 0
\end{equation}
with the equality again holding only when $\bm u' = \bm b' = 0$. Thus, based on Eq. \eqref{eq:M} the quantity $\mathcal{M}$ will always decay until the state $\bm u' = \bm b' = 0$ is reached, which is the only case for which $\partial_t \mathcal{M}=\mathcal{M}=0$.
Therefore, all perturbations to the basic state $\bm B_0$ will decay to zero independently of the value of $\nu,\,\mu > 0$. 

\begin{acknowledgements}
The authors acknowledge interesting and useful discussions
with Stephan Fauve. V.D. acknowledges the financial support from EU-funded Marie Curie Actions--Intra-European Fellowships (FP7-PEOPLE-2011-IEF, MHDTURB, Project No. 299973). The computations were performed using the HPC resources from GENCI-TGCC-CURIE (Project No. x2014056421).
\end{acknowledgements}
\bibliography{references}
\end{document}